# Coherent Detection of Discrete Variable Quantum Key Distribution using Homodyne Technique

Ayesha Jamal[1], Tahir Malik[2], Muhammad Kamran[3], Muhammad Fahim Ul Haque[2], and Muhammad Mubashir Khan[3]

[1]Department of Computer Science and Information Technology, NED University of Engineering and Technology, Karachi, Pakistan

[2]Department of Telecommunication Engineering, Faculty of Engineering, NED University of Engineering and Technology, Karachi, Pakistan

[3]Department of Computer Science and Information Technology, Faculty of Information Sciences & Humanities, NED University of Engineering and Technology, Karachi, Pakistan

**In Discrete Variable Quantum Key Distribution (DV-QKD), homodyne detection method is frequently employed for its simplicity in use, effectiveness in terms of error correction, and suitability with contemporary optical communication systems. Being a coherent detection method, it relies on a local oscillator whose frequency is matched to that of the transmitted carrier's signal. In this paper we evaluate a Free Space Optical (FSO) DV-QKD system based on the KMB09 protocol using Homodyne detection under random phase fluctuation and depolarizing noise error. We present simulation results for System Efficiency and Quantum Bit Error Rate (QBER) for the proposed model. An obtained efficiency ($\approx 25\%$) for our proposed DV-QKD system model shows that under atmospheric turbulence and noise effect, it is inline with the available analytical results. However, the inclusion of random phase fluctuation and noise led to higher-than-normal QBER which is anticipated in a real-world scenario.**

It is commonly known that complex Quantum Mechanics principles ensure the security of Quantum Key Distribution (QKD) by assuring a secure key transfer among both the parties involved [1–4]. These QKD protocols are often assessed according to their design uniqueness, accuracy, safety, and their ease of use [5]. Bennett & Brassard proposed the first, most fundamental and groundbreaking QKD protocol, known as BB84. It is based on Heisenberg's Uncertainty Principle and uses four particle states to encrypt and decrypt the information [6, 7]. Bennett's other protocol, the B92, is a simplified version of the BB84 protocol where bits are encoded in photons using just two non-orthogonal states [5, 8]. In contrast, the KMB09 protocol proposed by Khan, Murphy, and Beige in 2009 is a high dimensional QKD protocol which improves BB84 to be robust against Photon-Number-Splitting (PNS) attacks [9, 10]. Depending on how classical data is encrypted and decrypted, the majority of these quantum key distribution (QKD) protocols fall into one of two categories: discrete variable QKD (DV-QKD) or continuous variable QKD (CV-QKD) [11]. The DV-QKD system utilizes the no-cloning theorem and the theory on the indistinguishability of random quantum states to provide unconditional security [12]. Contrarily, the CV-QKD uses the uncertainty principle to argue that it is impossible to simultaneously measure the in-phase and quadrature elements of a coherent state with absolute precision [13, 14].

The homodyne detection method is widely employed in DV-QKD because of its simplicity, efficiency in error correction, and compatibility with modern optical communication protocols [15, 16]. Such systems secure the key information using different characteristics of a photon for encoding, such as polarization or the phase, then decode it using detection methods [17]. Homodyne detection is a specific kind of coherent detection technique which includes mixing of carrier signal with the reference signal (typically a local oscillator) to extract the key information from carrier signal which has the same frequency as the local oscillator [18–20]. Heterodyne detection is another type of coherent detection method where the carrier signal and local oscillator have different frequencies. [21–23]. This technique is an improvement on the traditional homodyne system and allows for the simultaneous detection of two field quadratures in one measurement [24]. Although there are clear advantages to the heterodyne technique, however the complexity of the equipment required for heterodyne detection can lead to greater implementation costs [25]. We focus our study on examining and analysing the homodyne method as it is a comparatively simpler method and offers ease of compatible with optical communication technologies [15].

The homodyne DV-QKD systems can be employed over wired (optical fibre) communication links as well as wireless (free-space optical (FSO)) communication links [26]. FSO is the term for line-of-sight (LoS) optical beam transmission over the atmosphere, which is intrinsically more adaptable and less expensive to deploy compared to optical fibre [27]. However, its overall performance is restricted by atmospheric factors like turbulence, absorption, and scattering phenomena, in addition to the aiming errors caused by an improper alignment among the sender and the receiver [28]. These effects reduce the carrier signal's quality, decreasing the performance of quantum bit error rate (QBER), or even leading to link failure entirely [27, 29].

Here we have evaluated and simulated the operation of a DV-QKD system with special attention to the KMB09 protocol, which assures that the security of the QKD system as a whole is contingent on the minimal ITER rate when an eavesdropper is present [9]. While we combine the DV-QKD approach with the homodyne detection, to make the model more realistic we additionally incorporate transmission losses brought on by air turbulence, random phase change and depolarizing noise to the system model. We contrast the outcomes of our model with the performance of a ideal KMB09 system without losses. Atmospheric turbulence is modelled using the von Kármán Spectrum model. It is extensively employed for fluctuating temperatures and varying velocity in the literature on random turbulence and propagation of electromagnetic waves in random media [30, 31]. The depolarizing noise effect is also added to this channel that causes the qubits in free space to deviate from their original state with the error probability $\rho$ due to random interactions [32, 33]. We have simulated this model at a fixed depolarizing angle of $\pi/4$ but also observed the outcomes of different depolarizing angles to gain deeper insights of DV-QKD model of KMB09. Finally, we have determined the QBER and efficiency of our proposed DV-QKD system for up to 1000 iterations of data.

The format of the paper is as follows: Section 2 presents the theoretical background of the evolution of DV-QKD in homodyne detection system. In Section 3, we represent our simulation model of DV-QKD-based homodyne detection. Section 4 provides the results of our simulations along with the comparison between them, and the conclusion is outlined in Section 5.

# 1 Theoretical Background

## 1.1 Discrete Variable Quantum Key Distribution (DV-QKD)

In DV-QKD, the information is often encoded into discrete optical properties of an individual photon, such as polarization, while decoding is usually carried out by a single photon detector [11, 34]. While multiple degrees of freedom are possible for a photon, these individual degrees of freedom can be utilized to store quantum information in DV-QKD [18, 35].

Long-distance DV-QKD methods have been practically evaluated and they offer more sophisticated security proofs that are considered for system errors and the consequences of finite information size [36–38]. BB84, SARG04, B92, along with the synchronisation protocols are some common DV-QKD protocols [11, 39, 40].

In addition to DV-QKD, the resistance of the KMB09 protocol to two real-world attack models has already been established in literature [41]. The polarisation-based DV-QKD approach which integrates the usage of Phase-Modulators for generating polarization states and alternating between the basis with a polarisation diversity homodyne detection technique implementing BB84 QKD protocol is one of the most recent works to take this into consideration [18]. In this paper, we consider a free space optical (FSO) channel using the KMB09 protocol with Homodyne detection at the receiver. DV-QKD along with the homodyne detection is simple to use, effective in reducing errors and compatible with current optical communication technology [15].

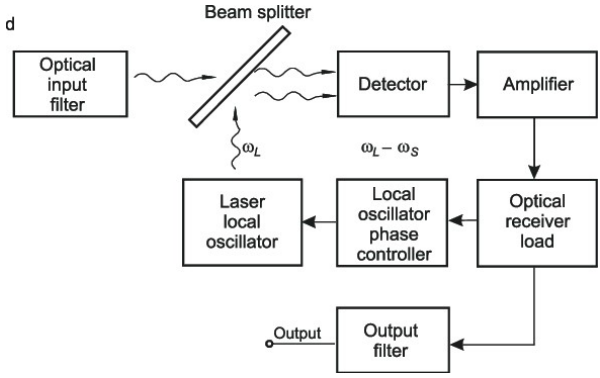

Figure 1: Block schematic of an optical receiver for homodyne detection.

## 1.2 Homodyne Detection Technique

Homodyne detection is a coherent detection technique that retrieves data from a light modulated signal by combining an identical-frequency local oscillator signal with the incoming signal (Fig. 1). The relevant information is then extracted from the resultant mixed signal by passing it through a low-pass filter [42–45]. The optical power of a coherent detector can be represented as [46],

$$P(t) = P_S(t) + P_{LO}(t) + 2\sqrt{P_S + P_{LO}} \cdot \cos[\omega_{IF}(t) + (\Phi_S(t) - \Phi_{LO}(t))] \quad (1)$$

Here, $P_S(t)$ is the optical power of received signal and can be expressed as $P_S(t) = |A_S(t)|^2$. $P_{LO}(t)$ is the optical power of Local Oscillator (LO) which can be expressed as $P_{LO}(t) = |A_{LO}(t)|^2$. 'A' represents the Amplitude. $\omega_{IF}(t)$ is the carrier frequency of the LO and $\Phi_S(t)$ and $\Phi_{LO}(t)$ are the phases of received signal and the LO respectively. The simplicity in Homodyne detection lies in matching $\omega_S$ to $\omega_{LO}$ thus having $\omega_{IF} = 0$. Assuming $\Phi_S(t) = \Phi_{LO}(t)$, we can neglect the phase difference and simplify eq. 1 as,

$$P(t) = P_S(t) + P_{LO}(t) + 2\sqrt{P_S + P_{LO}} \quad (2)$$

We must have a complete control on $P_{LO}$ so if $P_S$ is compromised in any condition, we can change the power of Local Oscillator accordingly. Since $P_S$ typically has a very low value compared to $P_{LO}$ (i.e., $P_{LO} \gg P_S$), $P_S$ can usually be disregarded, and $P_S + P_{LO} = P_{LO}$. However for our simulation, we have not neglected the value of $P_S$.

## 2 DV-QKD Based Coherent Detection Model

In this section we propose our novel DV-QKD system model, based on KMB09, using homodyne detection approach in an FSO Channel. The classical QKD model comprises of a sender i.e. Alice and receiver Bob. Alice prepares the KMB09 states and transmits them over the channel in this instance. The quadratures are then measured at random by Bob. Figure 2 shows the visual interpretation of our DV-QKD system based homodyne detection technique, which has three components: the quantum transmitter (Alice), the FSO channel and the quantum receiver (Bob).

### 2.1 Polarization State Generation

Alice generates the KMB09 polarisation states in one of the two bases $e_A$ or $f_A$ (A = 1, 2... N) by randomly switching among them (where all states of $e$ denote 0, and all states of $f$ denote to 1). Alice then combines the states with optical pulses (while considering it to be a truly single-photon source) followed by a Phase Modulator with a phase shift of $\pi/4$ and an Amplitude Modulator (the Mach Zehnder Modulator) and transmits them to Bob over a free space channel. The encoding method employed by Alice and Bob is displayed in Table 1.

Although our simulation only considers N = 2 (where N is the number of dimensions), it is possible to change the polarization state dimensions to any higher level (higher dimension) in order to increase security, which is still another crucial aspect of KMB09 protocol [9]. Moreover, the HD-QKD (i.e. KMB09) protocol's resistance to two actual attack models, the beam splitter (BS) and the photon number splitting (PNS) attack, with the incorporation of decoy-state scheme has previously been demonstrated in literature [41].

### 2.2 Transmission of states over a Free Space Channel

Polarization states prepared by Alice are then transmitted over a 1000 meter FSO channel. To improve the accuracy of our simulation model both the depolarizing noise effect and atmospheric turbulence are introduced to the FSO channel. The depolarizing noise effect causes the qubits in the free space to vary from their initial state due to random interactions at a fixed angle of $\pi/4$. Whereas the atmospheric turbulence causes changes to the phase of the receiving signal. The Von-Karman spectrum for the random phase is applied to our model [4]:

$$\Phi(\kappa) = \frac{[C(\alpha)r c^{-\alpha}]}{(\kappa^2 + \kappa_0^2)^{(1+\alpha/2)}} \exp\left(-\frac{\kappa^2}{\kappa_m^2}\right), \quad (3)$$

$$C(\alpha) = \frac{\alpha \cdot 2^{(\alpha-2)} \Gamma\left(1 + \frac{\alpha}{2}\right)}{\pi \cdot \Gamma\left(1 - \frac{\alpha}{2}\right)} \quad (4)$$

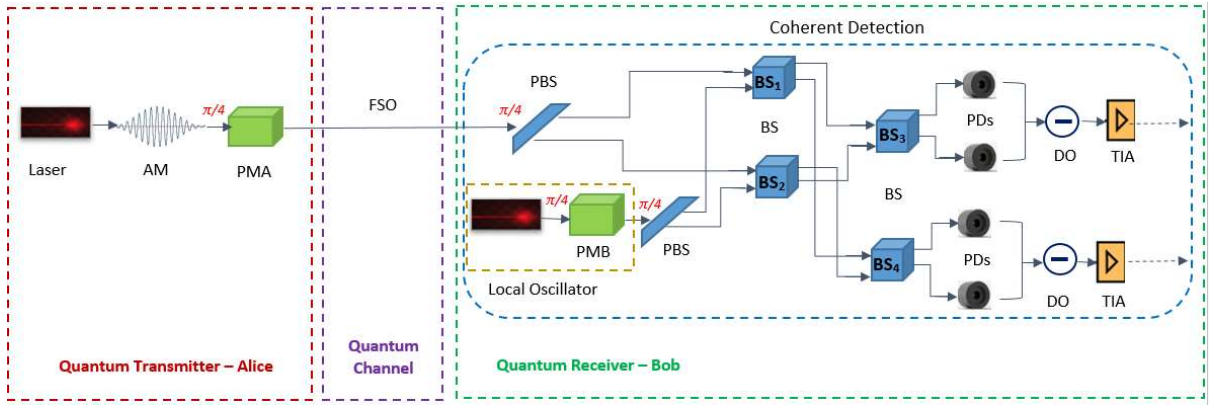

Figure 2: Visual Interpretation of our Discrete-Variable Quantum Key Distribution based on homodyne Detection model. Here [AM] is an Amplitude Modulator, [$PM_A$] and [$PM_B$] are Alice's and Bob's Phase Modulators respectively. [FSO] is the Free Space Channel, [PBS] denotes the Polarization Beam Spitters, [BS] are Beam Splitters, [PDs] the Photo-Detectors, [DO] the Difference Operators and [TIA] denotes the Trans-Impedance Amplifiers.

Table 1: Encoding Scheme for KMB09 Protocol

| Alice announced the Index | $|e_1\rangle$ | $|e_2\rangle$ | ... | $|e_N\rangle$ | $|f_1\rangle$ | $|f_2\rangle$ | ... | $|f_N\rangle$ |
|---|---|---|---|---|---|---|---|---|
| 1 | × | 1 | ... | 1 | × | 0 | ... | 0 |
| 2 | 1 | × | ... | 1 | 0 | × | ... | 0 |
| ... | ... | ... | ... | ... | ... | ... | ... | ... |
| N | 1 | 1 | ... | × | 0 | 0 | ... | × |

Where $\kappa$ is the spatial frequency that varies at each event, $\Gamma$ is the gamma-function, $\alpha$ is a constant, $\kappa_m = 2\pi/lo$, $\kappa_o = 2\pi/Lo$, where lo and Lo are inner and outer scales of turbulence respectively. $C(\alpha)$ is used to assess the magnitude of the phase changes brought on by turbulence. This random phase change caused by turbulence alters the entire phase of the incoming signal and as a result, the $\Phi_S(t)$ in our equation (1) will now become $(\Phi_S(t) \pm \Phi(\kappa))$, also in a real-world scenario, the $\Phi_S(t)$ and $\Phi_{LO}(t)$ are not equal, hence the $\cos[\omega_{IF}(t) + (\Phi_S(t) - \Phi_{LO}(t))]$ part from equation (1) is not equals to 1 and will be considered in the detection.

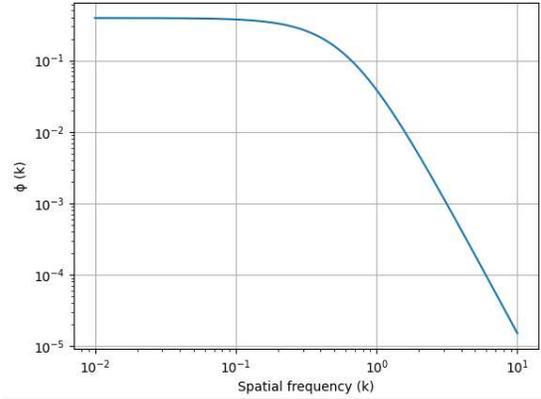

Figure 3: The Von-Karman power spectrum model for the random phase, where lo = 1cm, Lo = 10m, $\alpha$ = 5/3 over the distance of L = 1000 meters with a wavelength $\lambda$ = 1500nm.

## 2.3 Homodyne Detection at Receiving End

Typically, Bob switches between the two bases $e$ and $f$ at random to measure the incoming photon states from Alice and notes the results of his measurement. But in homodyne detection, Bob prepares his own set of states by alternating between $e_B$ and $f_B$ (B = 1, 2,...N). The states prepared by Alice and Bob, separately, are then combined with the help of a mixer (consists of 50/50 Beam Splitters (BS)) that send this combined state for the detection to Photodetector (PD). For the sake of simplicity we have kept the PBS, BS, and PD ideal, with no noise and phase errors. Following the transmission and detection of states, Alice communicates to Bob over a public channel to share the index of each state that she has transmitted. Bob compares his results to those that Alice made public and declares all the mismatches. In accordance with Table 1, Alice and Bob interpret their results and eliminate any instances where they do not have a shared key value. Returning to equation (1), after the effect of random phase fluctuation on the phase of our incoming qubits, equation (1) will now be as follows:

$$P(t) = P_S + P_{LO} + 2\sqrt{P_S + P_{LO}} \cdot \cos[\omega_{IF} + (\Phi_S(t) \pm \Phi(\kappa)) - \Phi_{LO}] \quad (5)$$

The homodyne receiver's total power P(t) now includes an additional phase $\Phi(\kappa)$ which ultimately contributes to error rate resulting in an increased QBER.

## 3 Simulation Results and Discussion

In this section, we will discuss our simulation results of Quantum Bit Error Rate (QBER) and system efficiency at each iteration. Depolarizing noise and turbulence effects are included in our model and the results are contrasted for iterations ranging from 250 to 1000. Figure 4 presents the QBER results of our system for different angles (0, $\pi/8$, $\pi/4$, $3\pi/8$ and $\pi/2$) of the depolarizing noise model, with and without turbulence effect. The QBER calculations from our simulation results are shown in figure 5 below, for upto 1000 iterations of data.

From figure 4, it is evident that QBER has increased as a result of turbulence. We observe that for $\vartheta = 0$, the QBER attained is 0.12 which is purely due to turbulence effects. With increasing value of $\vartheta$, we see an increase in QBER. The maximum change in QBER due to turbulence is observed at $\vartheta = 3\pi/8$, whereas the maximum value of QBER is attained at a $\vartheta$ of $\pi/2$.

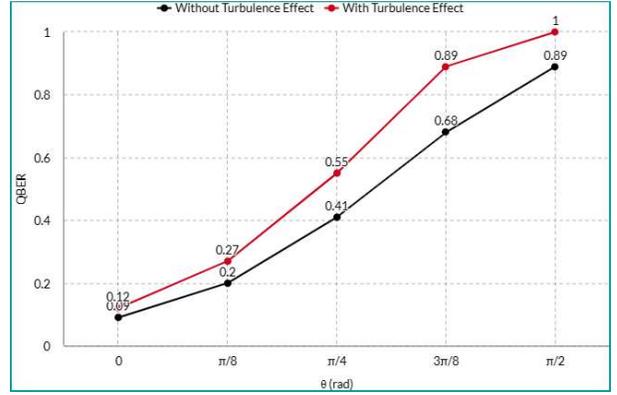

Figure 4: Effect of Depolarizing Noise Model on QBER for our simulated system of KMB09 in the absence & presence of Turbulence Effect ($\Phi(\kappa)$).

The QBER comparison of BB84 and KMB09 under the channel noise effect is further illustrated in Table 2. For an angle of $\pi/4$ we observe from Table 2 that the QBER obtained for our system model is similar to that obtained in [48] which uses BB84 QKD protocol over a similar type of noise channel model i.e. the collective-rotation noise channel. Where, $\vartheta$ represents the rotational angle at which the applied noise effect rotates the incoming polarization states from Alice to Bob. The Table 2 shows that the QBER increases by approximately 10% under conditions of noise and turbulence.

Table 2: Comparison of BB84 and KMB09 for similar noise channel models

| Parameters | BB84 | Our Simulated Model of KMB09 |
|---|---|---|
| $\vartheta$ | $\pi/4$ | $\pi/4$ |
| QBER | 0.50 | ≤ 0.55 |
| Eavesdropper | Absent | Absent |

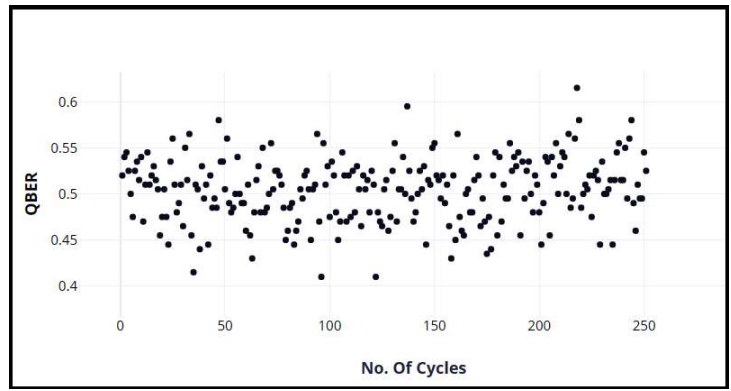

(a) QBER for 250 No. of Iterations

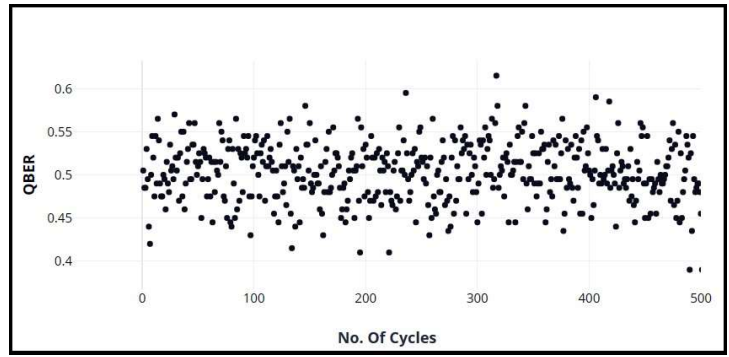

(b) QBER for 500 No. of Iterations

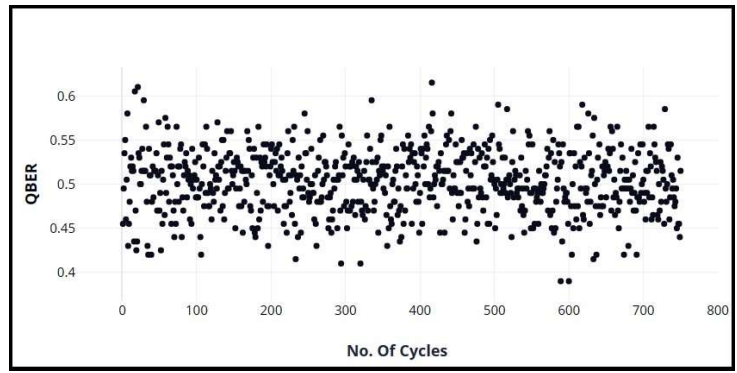

(c) QBER for 750 No. of Iterations

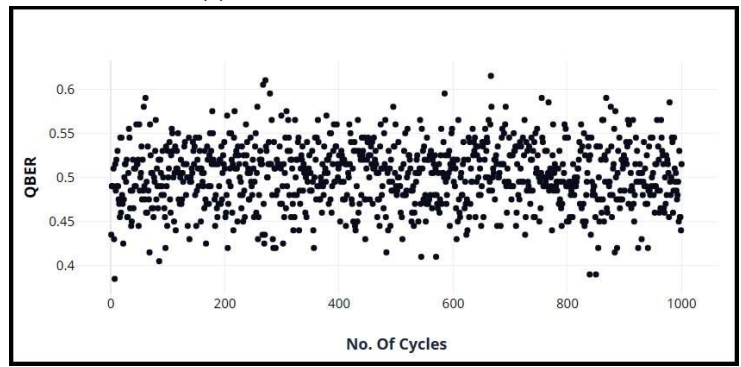

(d) QBER for 1000 No. of Iterations

Figure 5: Simulation results of QBER for our DV-QKD homodyne detection model, where $P_S$ = 3mV, $P_{LO}$ = 8mV and $\lambda$ = 1500nm, over the distance of 1 km.

We have also determined the efficiency of our system, which is a widely recognised system

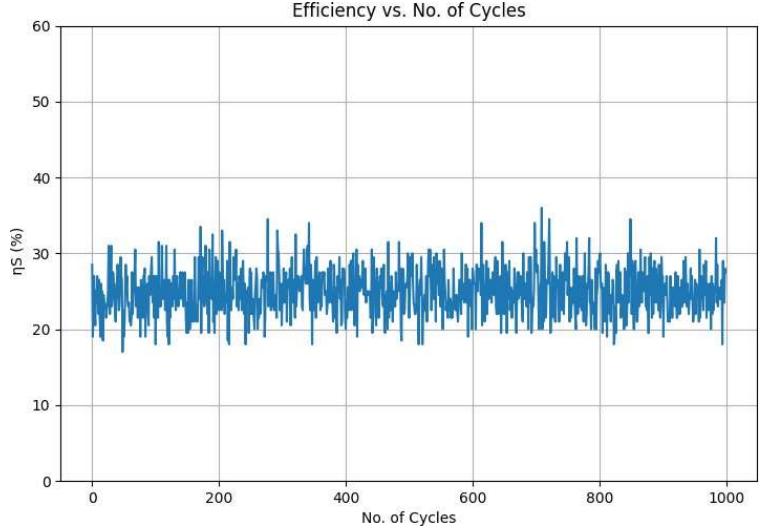

Figure 6: Average Efficiency for our DV-QKD Homodyne Detection System for 1000 iterations.

statistic for a QKD system [49, 50]:

$$\eta_A = \frac{N-1}{2N} \qquad (6)$$

Where $\eta_A$ is an analytical efficiency and N represents the number of dimensions [51].

For determining the system efficiency ($\eta_S$), 1000 iterations are performed and $\eta_S$ is noted against each iteration. With the increasing number of iterations, figure 6 shows that $\eta_S$ converges at the value of 0.25 (25%) for two dimensions (N = 2). This demonstrates that our simulated KMB09 DV-QKD system's efficiency is equivalent to that of the analytical model of KMB09.

## 4 Conclusion

The homodyne detection method is frequently implemented in discrete variable quantum key distribution (DV-QKD) to improve security and optimize system performance. In this paper we have demonstrated the DV-QKD homodyne system that takes KMB09 polarization states over an FSO channel. The FSO channel includes depolarizing noise along with the Von-Karman power spectrum model for random phase fluctuation i.e. turbulence. Our results show that QBER increases in severity with increasing noise and turbulence thus impacting the usability of the KMB09. We also show that the efficiency for our system is comparable to that of the KMB09 analytical model.

## Acknowledgement

We acknowledge funding from the Sindh Higher Education Commission, Pakistan under the Sindh Research Support Program (SRSP) at NED University of Engineering and Technology, Karachi.

## References

[1] F. Grünenfelder, A. Boaron, D. Rusca, A. Martin, and H. Zbinden. "Simple and high-speed polarization-based QKD". In: *Applied Physics Letters* 112.5 (2018).

[2] J. Barrett, L. Hardy, and A. Kent. "No signaling and quantum key distribution". In: *Physical review letters* 95.1 (2005), p. 010503.

[3] S. Pironio, A. Acín, N. Brunner, N. Gisin, S. Massar, and V. Scarani. "Device-independent quantum key distribution secure against collective attacks". In: *New Journal of Physics* 11.4 (2009), p. 045021.

[4] P. Wallden, V. Dunjko, A. Kent, and E. Andersson. "Quantum digital signatures with quantum-key-distribution components". In: *Physical Review A* 91.4 (2015), p. 042304.

[5] A. A. Abushgra. "Variations of QKD protocols based on conventional system measurements: A literature review". In: *Cryptography* 6.1 (2022), p. 12.

[6] M. H. Saeed, H. Sattar, M. H. Durad, and Z. Haider. "Implementation of QKD BB84 Protocol in Qiskit". In: *2022 19th International Bhurban Conference on Applied Sciences and Technology (IBCAST)*. IEEE. 2022, pp. 689–695.

[7] M. Elboukhari, M. Azizi, and A. Azizi. "Quantum Key Distribution Protocols: A Survey." In: *International Journal of Universal Computer Science* 1.2 (2010).

[8] C. H. Bennett. "Quantum cryptography using any two nonorthogonal states". In: *Physical review letters* 68.21 (1992), p. 3121.

[9] M. M. Khan, M. Murphy, and A. Beige. "High error-rate quantum key distribution for long-distance communication". In: *New Journal of Physics* 11.6 (2009), p. 063043.

[10] G. Dai. "Formal Verification for KMB09 Protocol". In: *International Journal of Theoretical Physics* 58 (2019), pp. 3651–3657.

[11] I. W. Primaatmaja, C. C. Liang, G. Zhang, J. Y. Haw, C. Wang, and C. C.-W. Lim. "Discrete-variable quantum key distribution with homodyne detection". In: *Quantum* 6 (2022), p. 613.

[12] I. B. Djordjevic. "On the discretized Gaussian modulation (DGM)-based continuous variable-QKD". In: *IEEE Access* 7 (2019), pp. 65342–65346.

[13] I. B. Djordjevic. "Hybrid QKD protocol outperforming both DV-and CV-QKD protocols". In: *IEEE Photonics Journal* 12.1 (2019), pp. 1–8.

[14] S. Wijesekera, X. Huang, and D. Sharma. "Quantum cryptography based key distribution in Wi-Fi networks-Protocol modifications in IEEE 802.11". In: *International Conference on Software and Data Technologies*. Vol. 2. SCITEPRESS. 2010, pp. 146–151.

[15] W.-B. Liu, C.-L. Li, Y.-M. Xie, C.-X. Weng, J. Gu, X.-Y. Cao, Y.-S. Lu, B.-H. Li, H.-L. Yin, and Z.-B. Chen. "Homodyne detection quadrature phase shift keying continuous-variable quantum key distribution with high excess noise tolerance". In: *PRX Quantum* 2.4 (2021), p. 040334.

[16] L. Oesterling, D. Hayford, and G. Friend. "Comparison of commercial and next generation quantum key distribution: Technologies for secure communication of information". In: *2012 IEEE Conference on Technologies for Homeland Security (HST)*. IEEE. 2012, pp. 156–161.

[17] A. Duplinskiy, V. Ustimchik, A. Kanapin, V. Kurochkin, and Y. Kurochkin. "Low loss QKD optical scheme for fast polarization encoding". In: *Optics express* 25.23 (2017), pp. 28886–28897.

[18] M. F. Ramos, A. N. Pinto, and N. A. Silva. "Polarization based discrete variables quantum key distribution via conjugated homodyne detection". In: *Scientific Reports* 12.1 (2022), p. 6135.

[19] S. Yang, T. Xing, C. Ke, J. Liang, and X. Ke. "Effect of Wavefront Distortion on the Performance of Coherent Detection Systems: Theoretical Analysis and Experimental Research". In: *Photonics*. Vol. 10. 5. MDPI. 2023, p. 493.

[20] R. Zhang, L. Li, Z. Zhao, G. Xie, G. Milione, H. Song, P. Liao, C. Liu, H. Song, K. Pang, et al. "Coherent optical wireless communication link employing orbital angular momentum multiplexing in a ballistic and diffusive scattering medium". In: *Optics Letters* 44.3 (2019), pp. 691–694.

[21] S. Kleis. "System Approaches for a Practical Implementation of Continuous-Variable Quantum Key Distribution Using Coherent Heterodyne Detection". PhD thesis. Universitätsbibliothek der HSU/UniBwH, 2022.

[22] X.-C. Ma, S.-H. Sun, M.-S. Jiang, and L.-M. Liang. "Wavelength attack on practical continuous-variable quantum-key-distribution system with a heterodyne protocol". In: *Physical Review A* 87.5 (2013), p. 052309.

[23] R. K. Goncharov, A. Zinovev, F. D. Kiselev, and E. O. Samsonov. "Heterodyne-based subcarrier wave quantum cryptography under the chromatic dispersion impact". In: *: ,* 12.2 (2021), pp. 161–166.

[24] R. Goncharov, I. Vorontsova, D. Kirichenko, I. Filipov, I. Adam, V. Chistiakov, S. Smirnov, B. Nasedkin, B. Pervushin, D. Kargina, et al. "The rationale for the optimal continuous-variable quantum key distribution protocol". In: *Optics* 3.4 (2022), pp. 338–351.

[25] H. P. Yuen and J. H. Shapiro. "Quantum statistics of homodyne and heterodyne detection". In: *Coherence and Quantum Optics IV: Proceedings of the Fourth Rochester Conference on Coherence and Quantum Optics held at the University of Rochester, June 8–10, 1977*. Springer. 1978, pp. 719–727.

[26] P. V. Trinh, A. T. Pham, A. Carrasco-Casado, and M. Toyoshima. "Quantum key distribution over FSO: Current development and future perspectives". In: *2018 Progress in Electromagnetics Research Symposium (PIERS-Toyama)*. IEEE. 2018, pp. 1672–1679.

[27] M. A. Khalighi and M. Uysal. "Survey on free space optical communication: A communication theory perspective". In: *IEEE communications surveys & tutorials* 16.4 (2014), pp. 2231–2258.

[28] N. Alshaer, M. E. Nasr, and T. Ismail. "Hybrid MPPM-BB84 quantum key distribution over FSO channel considering atmospheric turbulence and pointing errors". In: *IEEE Photonics Journal* 13.6 (2021), pp. 1–9.

[29] H. Kaushal and G. Kaddoum. "Optical communication in space: Challenges and mitigation techniques". In: *IEEE communications surveys & tutorials* 19.1 (2016), pp. 57–96.

[30] G. Goedecke, V. E. Ostashev, D. K. Wilson, and H. J. Auvermann. "Quasi-wavelet model of von Kármán spectrum of turbulent velocity fluctuations". In: *Boundary-layer meteorology* 112.1 (2004), pp. 33–56.

[31] F. L. Dos Santos, L. Botero, C. Venner, and L. D. de Santana. "Modelling the dissipation range of von Kármán turbulence spectrum". In: *AIAA Aviation 2021 Forum*. 2021, p. 2292.

[32] H. Qiao and X.-y. Chen. "Simulation of BB84 Quantum Key Distribution in depolarizing channel". In: *Proceedings of 14th Youth Conference on Communication*. 2009, pp. 123–129.

[33] X. Xu and X. Chen. "Simulating B92 Protocol in Depolarizing Channel". In: *2010 6th International Conference on Wireless Communications Networking and Mobile Computing (WiCOM)*. IEEE. 2010, pp. 1–3.

[34] S. Ghorai, P. Grangier, E. Diamanti, and A. Leverrier. "Asymptotic security of continuous-variable quantum key distribution with a discrete modulation". In: *Physical Review X* 9.2 (2019), p. 021059.

[35] M. Lasota, R. Filip, and V. C. Usenko. "Robustness of quantum key distribution with discrete and continuous variables to channel noise". In: *Physical Review A* 95.6 (2017), p. 062312.

[36] A. Boaron, G. Boso, D. Rusca, C. Vulliez, C. Autebert, M. Caloz, M. Perrenoud, G. Gras, F. Bussières, M.-J. Li, et al. "Secure quantum key distribution over 421 km of optical fiber". In: *Physical review letters* 121.19 (2018), p. 190502.

[37] K. Takemoto, Y. Nambu, T. Miyazawa, Y. Sakuma, T. Yamamoto, S. Yorozu, and Y. Arakawa. "Quantum key distribution over 120 km using ultrahigh purity single-photon source and superconducting single-photon detectors". In: *Scientific reports* 5.1 (2015), p. 14383.

[38] N. Lütkenhaus. "Security against individual attacks for realistic quantum key distribution". In: *Physical Review A* 61.5 (2000), p. 052304.

[39] D. Stucki, N. Brunner, N. Gisin, V. Scarani, and H. Zbinden. "Fast and simple one-way quantum key distribution". In: *Applied Physics Letters* 87.19 (2005).

[40] K. Inoue, E. Waks, and Y. Yamamoto. "Differential phase shift quantum key distribution". In: *Physical review letters* 89.3 (2002), p. 037902.

[41] M. Kamran, M. M. Khan, and T. Malik. "Decoy state HD QKD system for secure optical communication". In: *2021 International Conference on Cyber Warfare and Security (ICCWS)*. IEEE. 2021, pp. 87–92.

[42] T. Hirano, H. Yamanaka, M. Ashikaga, T. Konishi, and R. Namiki. "Quantum cryptography using pulsed homodyne detection". In: *Physical review A* 68.4 (2003), p. 042331.

[43] J. Lodewyck, M. Bloch, R. García-Patrón, S. Fossier, E. Karpov, E. Diamanti, T. Debuisschert, N. J. Cerf, R. Tualle-Brouri, S. W. McLaughlin, et al. "Quantum key distribution device with coherent states". In: *Quantum Communications Realized*. Vol. 6780. SPIE. 2007, pp. 150–163.

[44] B. Qi, L.-L. Huang, L. Qian, and H.-K. Lo. "Experimental study on the Gaussian-modulated coherent-state quantum key distribution over standard telecommunication fibers". In: *Physical Review A* 76.5 (2007), p. 052323.

[45] S. Fossier, E. Diamanti, T. Debuisschert, A. Villing, R. Tualle-Brouri, and P. Grangier. "Field test of a continuous-variable quantum key distribution prototype". In: *New Journal of Physics* 11.4 (2009), p. 045023.

[46] G. Keiser. *Optical Fiber Communications*. McGraw-Hill international editions. McGraw-Hill, 2010.

[47] I. A. Adam, D. A. Yashin, D. A. Kargina, and B. A. Nasedkin. "Comparison of Gaussian and vortex beams in free-space QKD with phase encoding in turbulent atmosphere". In: *: , ,* 13.4 (2022), pp. 392–403.

[48] M. Mafu, C. Sekga, and M. Senekane. "Security of Bennett–Brassard 1984 quantum-key distribution under a collective-rotation noise channel". In: *Photonics*. Vol. 9. 12. MDPI. 2022, p. 941.

[49] T. Zhou, J. Shen, X. Li, C. Wang, and J. Shen. "Quantum cryptography for the future internet and the security analysis". In: *Security and Communication Networks* 2018 (2018), pp. 1–7.

[50] J. Wang, Q. Zhang, and C.-j. Tang. "Quantum key distribution protocols using entangled state". In: *2006 International Conference on Computational Intelligence and Security*. Vol. 2. IEEE. 2006, pp. 1355–1358.

[51] M. Kamran, D. KHAN, T. Malik, and A. Arfeen. "Quantum key distribution over free space optic (FSO) channel using higher order Gaussian beam spatial modes". In: *Turkish Journal of Electrical Engineering and Computer Sciences* 28.6 (2020), pp. 3335–3351.